
\input harvmac
\magnification=1200
\hoffset = .75 in
\voffset = .2 in
\hsize = 4.9 in
\vsize = 7.5 in

{\vbox{\centerline{\bf HIDDEN SYMMETRY APPROACH TO THE EXTENDED }}
\vskip2pt\centerline{\bf SYMMETRIES IN NONABELIAN FREE FERMIONIC}
\vskip2pt\centerline{\bf AND WZNW MODELS}}

\centerline{Raiko P. Zaikov\footnote{$^\star$}
{Supported by National Fondation for Science under
grants Ph-11/91-94 and Ph-318/93-94}}
\bigskip\centerline{Institute for Nuclear Research and Nuclear Energy}
\centerline{Boul. Tzarigradsko Chaussee 72, 1784 Sofia}
\centerline{e-mail: zaikov@bgearn.bitnet}

\vskip .3in

The hidden symmetry approach (HSA) together with the operator
product formalism (first applyied by Zamolodchikov)
\ref\rZA{Zamolodchikov, A. B, Theor.  Math. Phys., 63 (1985)
1205.} and the Drinfeld-Sokolov (D-S) reduction procedure to the
degree of freedom of a theory whose structure is based on a Lie
algebra \ref\rGD{Gel'fand, I. M and Dickey, L. A., {\it "A
Family of Hamiltonian Structures Connected with Integrable
Non-Linear Differential Equations"}, IPM preprint, AN SSSR,
Moskow 1978} \nref\rDS{Drinfeld, V. and Sokolov, V., J. Sov.
Math., 30 (1984) 1975}--
\ref\rFL{Fateev, V. A. and Lukyanov, S. L., Int. J. Mod. Phys.,
A3 (1988) 507 \semi Sov. Phys. JETP, 67 (1988) 447 \semi Sov.
Sci. Rev. A Phys., 15 (1990) 1.} are used for construction of
higher spin extension of the Virasoro algebra.  A complete list
of references can be found in the Bouwknegt and Schoutens Review
\ref\rBS{Bouknegt, P. and Schoutens, {"\cal W}-{\it Symmetry in
Conformal Field Theory}",  CERN preprint CERN-TH.6583/92, to be
published in Phys.  Reports}. We recall, that the HSA consist of
searching for an additional symmetry for some two-dimensional
conformal invariant action providing additional conserved
quatities. Moreover, the HSA allows us to gauge this extended
symmetry \ref\rHu{Hull, C. M., Phys. Lett., B240 (1990) 110
\semi\ Phys.  Lett., B259 (1991) 68\semi\ Nucl.  Phys., B353
(1991) 707 \semi\ Nucl. Phys., B364 (1991) 621.}
\nref\rBPR{Bergshoeff, E., Pope, C. N., Romans, L. J., Sezgin,
E., Shen, X. and Stelle, K. S., Phys. Lett., B243 (1990)
350.}--\ref\rNPV{Nissimov, E., Pacheva, S. and Vaysburg, I.,
Phys. Lett., B284 (1992) 321.}.

In the present talk the HSA is applyied to WZNW model and
nonabelian free fermionic model (NAFF). This allows us to check
the Witten's nonabelian bosonisation procedure \ref\rW{Witten,
E., Comm. Math. Phys., 92 (1984) 455.} on the higher spin
symmetry level. It is shown that both WZNW model and the NAFF
model admit higher spin extension of the semidirect product of
the Virasoro algebra and the affine algebra.  Despite
the fact, that both extended algebras are isomorphic, the higher
spin WZNW Noether currents form nonivariant space, while the
space of the NAFF currents is invariant. This shows that there
is some nonequivalence between the WZNW model and the NAFF model
on the higher spin symmetry level.

The higher spin conserved currents for both models are: %
\eqn\ea{{\cal V}^{2n}=\psi \partial _z^{2n+1}\psi , \qquad
\qquad \ V^n=tr\bigl(\partial g^{-1}\partial ^{n+1}g\bigr),}
\eqn\eb{{\cal J}^n_a=\psi t_a\partial _z^{n}\psi , \qquad
\qquad  J^n_a=tr\bigl(\partial g^{-1}\partial ^{n}gt_a\bigr),
}
where $(n=0,1,\dots ), \ V^0=T_{zz}, \ J^0_a=J_a$ and \ $z$ \ is
a complex variable. We restrict our considerations only with the
holomorphic components $\partial _{\bar z}V=0$ , having in mind
that the antiholomorphic components have a similar form and we
shall concerned  with the $SO(N)$ and $GL(N)$ models only.

It is easy to show that the currents \ea , \eb \ are consequence
of the symmetry of the action with respect to the following
transformations \ref\rRZ{R. P. Zaikov, {\it Extended Non-Abelian
Gauge Symmetries in Classical WZNW Model}, preprint
INRNETH/1/93; hep-th/9303087}, \ref\rZra{R. P. Zaikov, {\it
Extended nonabelian Simmetries for Free Fermionic Model},
preprint ICTP, IC/93/237, Trieste 1993, hep-th/9308016} (see
also \ref\rAA{Abidalla, E., Abidalla, M. C., Sotkov, G. and
Stanishkov, M., {\it Of "Critical Current Algebra"}, preprint
IFUSP/1027/93, Sao Paulo: hep-th/9302002.}):
\eqn\ec{\delta ^n\Phi =k_n(z)\partial ^{n+1}\Phi ,}
\eqn\ed{\widehat \delta ^n\Phi =\alpha ^a_n(z)\partial
^{n+1}\Phi t_a,}
where $k(z)$ and $\alpha (z)$ are arbitrary holomorphic
functions, $\Phi =(\psi ,g)$ and $t_a$ are the generators of
$GL(N)$ group in the fundamental representation. We note, that
the generators $t_a$ for $SO(N)$ models must be completed to the
generators of $GL(N)$ group.

We note, that the transformations \ec , \ed \ are only an
on-shell symmetry of the action. For the WZNW model there exist
nonlinear off-shell symmetry transformations also which,
however, do not form closed algebra in the affine sector.

It is easy to check that the infinitesimal transformations \ec ,
\ed \ satisfy the following Lie algebra:
\eqn\etr{›\delta ^m(k),
\delta ^n(h)!\Phi (x)=
\sum ^{\max(m+1,n+1)}_{r=0}\delta ^{m+n-r+1}
\Bigl(›h_n,k _m!^r_-\Bigr)\Phi (x)}
\eqn\eff{›\delta ^m(k),\widehat \delta ^n(\hat \beta )!\Phi (x)=
\sum ^{max (m+1,n)}_{r=0}\widehat \delta ^{m+n-r+1}(›\hat \beta
,k!^r_-)\Phi (x).}
\eqn\efd{\eqalign{›\widehat \delta ^m(\hat \alpha ),
\widehat \delta ^n(\hat \beta )!\Phi (x)=
\sum ^{max(m,n)}_{r=0}\biggl(& \widehat \delta
^{m+n-r}_c(f_{ab}^c›\beta ^b,\alpha ^a!^r_+t_c)+\widehat \delta
^{m+n-r}(d_{ab}^c›\beta ^b,\alpha ^a!^r_-t_c) \cr + & 2\delta
^{m+n-r}(›\beta ^a,\alpha _a!^r_-)
\biggr)\Phi (x),\cr}}
where the following notations are used
\eqn\efda{2\lbrack \beta _n,\alpha _m\rbrack ^r_{\pm }=
\pmatrix{ n \cr r \cr }\beta _n\partial ^r\alpha _n\pm
\pmatrix{ m \cr r \cr }\alpha _m\partial ^r\beta _n}
and it is taken into account that the binomial coefficients
\ $\pmatrix{m \cr r \cr }=0$ \ for \  $r>m$.
This algebra contains as subalgebras the Virasoro algebra, the
$W_{\infty }$ algebra, semi-direct product of the Virasoro and
the affine algebras. However, as whole the algebra
\etr -\efd \ is not a semi-direct product of the $W_{\infty
}$ and the higher spin affine algebra which follows from \efd .

It is strightforward to derive the transformation law for the
fermionic currents \ea  \ under the transformations \ec :
\eqn\ee{\delta ^m{\cal V}^n=\sum_{p=0}^{n+1}\pmatrix {n+1 \cr p
\cr } \partial ^pk_m{\cal V}^{m+n-p+1}-
k_m\sum _{q=0}^{m+1}(-)^{m-q}\pmatrix{ m+1 \cr q \cr }
\partial ^q{\cal V}^{m+n-q+2},}
which show that the currents ${\cal V}^n$ form an invariant
space with respect to the transformations \ec . On the same way
it can be show that the currents ${\cal V}^n, {\cal J}^n_a$ form
invariant space with respect to the transformations \ec \ and
\ed \ too. However, this is not the case for the corresponding
WZNW currents. Indeed, in this case we obtain:
\eqn\esh{\eqalign{
\delta ^mV^n(x) & =tr\bigl(\partial (\delta ^mg^{-1})\partial ^{n+1}g +
\partial g^{-1}\partial ^{n+1}(\delta ^{m+1}g)\bigr) \cr
& =-\partial \Bigl(k_mtr\bigl(g^{-1}\partial ^{m+1}gg^{-1}
\partial ^{n+1}g\bigr)\Bigr)+
k_mtr\bigl(g^{-1}\partial ^{m+1}gg^{-1}\partial ^{n+2}g\bigr) \cr
& +\sum_{r=0}^{m+1}\pmatrix{m+1 \cr r \cr }
\partial ^rk_mtr\bigl(\partial g^{-1}\partial ^{m+n-r+2}g\bigr),\cr }}
where we  use the transformation law for $g^{-1}$:
\eqn\esi{\delta ^mg^{-1}=-g^{-1}\delta ^mg g^{-1}=
-k_mg^{-1}\partial ^{m+1}gg^{-1},}
which satisfyies the same Lie algebra \etr .  In the case $m=1$
corresponding to the conformal transformations we find from \esh
\eqn\eshh{\delta ^0V^n(x)=(n+2)\partial k_0V^n(x)+
k_0\partial V^n+\sum _{r=2}^{n+1}\pmatrix {n+1 \cr r \cr }
\partial ^rk_0V^{n-r+1},}
which shows that only for $n=0$ we have a primary field
transformation law and for $n>0$ we have quasi-primary
transformation law.

We note, that in the general case $m>0$ in the formula \esh \
there appear the fourth order (with respect to \ $g$ \ ) terms
due to which the WZNW currents \ea \ and \eb \ do not form an
invariant space.  It can be shown that in the general case $m>0$
it is impossible to reduce these higher degree terms into the
bilinear ones only.  To check the latter statement we consider
the first term in the second line of \esh \ for the simplest
notrivial case $m=n=1$
\eqn\esj{tr\{g^{-1}\partial ^2gg^{-1}\partial
^2g\}= V^2-\partial V^1-tr\{g^{-1}\partial g\partial
g^{-1}\partial ^2g\}.}
It is impossible to express the last term in \esh \ as a linear
combination of the bilinear currents only. To clarify this
observation we return to the transformation law \ec \ (for $g$)
from which we find (for $k_m(x)=1$)
\eqn\esja{\eqalign{\delta ^m(\delta ^ng) &
=\delta ^m(gU^{n+1})=
\delta ^mgU^{n+1}+g\delta ^mU^{n+1} \cr
& =g(U^{m+1}U^{n+1}-U^{m+1}U^{n+1}+U^{m+n+2})=
\delta ^{m+n+1}g,
\cr }}
where
\eqn\esjb{\delta ^mU^{n}=
\bigl(U^{m+n+1}-U^{m+1}U^{n}\bigr)}
and
\eqn\eska{U^m=g^{-1}\partial ^mg, \qquad (m=0,1,\dots ).}
The second term in \esjb \ can be represented linearly with
respect to $U$ only if $m=1$. The formula \esja \ shows  that
the nonlinear term which appears in the current transformation
law
\esjb \ is canceld in the field transformation law. The latter
gives a possibility to close the algebra in linear realization
in the case when the corresponding currents space is
noninvariant.

In order to have an invariant current space, according to \esh \
we must extend the current space $\{V^m;m=0,1,\dots \}$  (which
we call basic current space) with additional nonlinear currents:
\eqn\esz{V^{n_1,n_2,\dots ,n_l}=tr\bigl(U^{n_1}U^{n_2}\dots
U^{n_l}\bigr), \qquad (l=1,2,\dots ).}
Taking into accaunt that $U^0=I$ \ and \ $U^1=j=g^{-1}\partial
g$ from \esz \ we obtain $V^{n_1,\dots ,n_r,0,\dots
,0}=V^{n_1,\dots ,n_r}$ if $(r=2,3,\dots )$ and
$V^n=V^{1,n+1,0,\dots ,0}$. It is strightforward to verify that
the extended currents space $\{V^{n_1,n_2,\dots ,n_l}\}$ is
invariant with respect to the $W_{\infty }$-transformations \ec
\eqn\esza{\eqalign{\delta ^mV^{n_1,n_2,\dots ,n_l}
& =\sum _{q=0}^{l-1}\Bigl\{-k_mV^{n_1,\dots ,n_q,m,n_{q+1},\dots
,n_l} \cr
& +\sum _{p=0}^{n_q+1}\pmatrix {n_q+1 \cr p \cr }\partial
^pk_mV^{n_1,\dots ,n_{q-1}n_q+m-p+1,n_{q+1},\dots
,n_l}\Bigr\}.\cr }}

To have an invariant current space with respect to the extended
affine transformations too, we include  symmetric isotopic
tensor currents $J^{n_1,n_2,\dots ,n_l}_{a_1,\dots ,a_p},
(l,p=1,2,\dots ; \ p<l)$. For example the explicit form of the
rank $2$ tensor currents is:
\eqn\eszb{J^{n_1,\dots ,n_l}_{a,b}=\sum _{q,p=0,q\ne
p}^ltr\Bigl(U^{n_1}\dots U^{n_p}t_a\dots U^{n_q}t_b\dots
U^{n_l}\Bigr).}
It is strightforward to check that the set of
all tensor currents form an invariant space.

The conformal spins for the currents \esz \ and \eszb \ are
$$
s_l=\sum _{q=0}^ln_q.
$$
The currents $V^{1,n_1,\dots ,n_p}$ are also Noether quantities
which correspond to the following nonlinear transformations:
\eqn\eszc{\delta ^{\{n\}}g=k_{\{n\}}gU^{n_1}\dots U^{n_l}.}
If $n_1=n_2=\dots =n_l=1$ these transformations coincide with
$w_{\infty }$ transformations \ $\delta ^lg=k_lgj^{l+1}$ \ and
the corresponding currents $v^l=tr(j^{l+2})$ form an invariant
space. In any other case the transformations \eszc \ close a
very wide algebra containing
\etr
\ as a subalgebra. In the case $n_r\ne 0,1$ for any $r=1,\dots
,p$  the currents \esz \ are not Noetherian. This is the
case for the currents \eszb \ too. We note, that the extended
current space \eszb \ is not invariant with respect to the
transformations \eszc \ and hence in order to retain this
invariance we must complete \eszb \ with currents including
derivatives of $U$ also. Consequently, this procedure give an
``explosion'' of the $W_{\infty }$ algebra.

The invariance of the extended currents space \eszb \ with
respect to the transformation \ec \ and \ed \ makes possible the
gauging of the symmetries considered here.

We note, that the above discussed nonivariances of the basic
(Noether) currents space with respect to their generating
transformations is a characteristic property of the WZNW model,
as well as of the principal chiral model. This is a consequence
of the transformation law for the field $g^{-1}$ \esi . This
phenomenon makes the difference between the $O(N)$
free-fermionic model (see \rZra ) and the corresponding WZNW
model which are completely equivalent on the ordinary symmetry
level \rW .

\listrefs
\bye